\documentstyle[epsfig,preprint,aps]{revtex}
\begin{document}
\draft         
\preprint{\vbox{\hbox{IFT--P.028/97}\hbox{FTUV/97-14}\hbox{IFIC/97-14}
\hbox{hep-ph/9703430}}}

\title{Discriminating New Physics Scenarios at NLC: \\
The Role of Polarization}
\author{E.\ M.\ Gregores$^1$, M.\ C.\ Gonzalez--Garcia$^{1,2}$,
and S.\ F.\ Novaes$^1$}
\address{$^1$Instituto de F\'{\i}sica Te\'orica, 
             Universidade Estadual Paulista \\   
             Rua Pamplona 145,
             01405--900 S\~ao Paulo, Brazil} 
\address{$^2$Instituto de F\'{\i}sica Corpuscular - IFIC/CSIC,
             Departament de F\'{\i}sica Te\`orica \\
             Universitat de Val\`encia, 46100 Burjassot, 
             Val\`encia, Spain}
\date{\today} 

\maketitle
\widetext
\begin{abstract} 
We explore the potential of the Next Linear Collider (NLC),
operating in the $e\gamma$ mode, to disentangle new physics
scenarios on single $W$ production. We study the effects related
with the exchange of composite fermion in the reaction $e\gamma
\to W \nu_e$, and compare with those arising from trilinear gauge
boson anomalous couplings. We stress the role played by the
initial state polarization to increase the reach of this machine
and to discriminate the possible origin of the new phenomena.
\end{abstract}

\section{Introduction}

The Standard Model (SM) of the electroweak interactions has
received a striking confirmation of its predictions after the
recent set of precise measurements made by CERN Large
Electron--Positron Collider (LEP) \cite{ewwg}. In particular,
the properties of the neutral weak boson and its couplings with
fermions were established with great precision. However, we still
lack the same confidence on other sectors of the SM, like the
self--couplings among the vector bosons, which is determined by
the non--abelian gauge structure of the theory. Moreover, the SM
does not furnish any reasonable explanation for the  replication
of fermionic families and their  pattern of mass.

In the search for an explanation for the fermionic generations,
we face theories where the known particles (leptons, quarks, and
gauge bosons) are composite \cite{comp}, and share common
constituents. In this case, the SM should be seen as the low
energy limit of a more fundamental theory, whose main feature
would be the existence of excited states, with mass below or of
the order of some large mass scale $\Lambda$.  

Searches for composite states have been carried out by several
collaborations of the CERN LEP collider \cite{lep}, and also from
DESY electron--proton collider, HERA \cite{hera}. Recent data
from LEP experiments have excluded excited spin--$\frac{1}{2}$
electrons with mass up to 80 GeV from the pair production search,
and up to 160 GeV from direct single production, for a scale of
compositeness $f/\Lambda > 0.7$ TeV$^{-1}$ (see below). HERA
experiments are able to exclude excited electrons with mass below
200 GeV for $f/\Lambda > 4.9$ TeV$^{-1}$. Bounds on excited
electron couplings have also been established through the
evaluation of radiative corrections to $Z^0$ width at one--loop
level \cite{nov96}. On the theoretical side, there have been
extensive studies on the possibility of unravelling the existence
of excited fermions in $pp$ \cite{pp,kuhn}, $e^+e^-$
\cite{kuhn,ele:pos,hag,lep:hera,e:gam}, and $ep$
\cite{hag,lep:hera} collisions at higher energies. 

As pointed out above, another possible source of deviation from
the SM predictions, that is still allowed by experimental
results, is the existence of anomalous vector boson
self--interactions. Our knowledge of the structure of the
trilinear couplings between gauge bosons remains rather poor
despite several experimental efforts \cite{anoc:exp}.  One of the
main goals of LEPII collider at CERN will be the study of the
reaction $e^+e^- \rightarrow W^+W^-$ which will furnish stronger
bounds on possible anomalous $W^+W^-\gamma$ and $W^+W^-Z$
interactions \cite{lepiiwg}. There exist in the literature
several theoretical studies on the probes for these anomalous
interactions on  future $e^+e^-$ \cite{ano:nlc,anoc:pheno} and
$pp$ \cite{ano:lhc} colliders. 

An important tool for the search of new physics will be the Next
Linear Collider (NLC), an $e^+e^-$ collider that will have a
center--of--mass energy of at least $500$ GeV with an integrated
luminosity around 50 fb$^{-1}$ \cite{nlc}.   At NLC, it will be
possible to obtain a high energy photon beam via the Compton
scattering of a laser off the electron beam
\cite{las0,laser:pol}.  The so called laser backscattering
mechanism will permit to obtain reactions initiated by either
$e^+e^-$, $e\gamma$, or $\gamma \gamma$ at NLC.

Of particular interest in the search for both excited fermions
and anomalous gauge boson self--interactions is the $e\gamma$
mode of NLC. Through this scattering it is possible, for instance,
to search for new excited charged leptons as $s$--channel
resonances \cite{e:gam}. Moreover, it has also been demonstrated
by several authors \cite{anoc:pheno} that single $W$ production
at the NLC operating on its $e\gamma$ mode is an ideal channel
for searching for deviations from the SM in the gauge
self--coupling sector. 

In this work we analyze the deviations from the SM predictions
for the reactions $e\gamma \rightarrow W\nu_e$  at
$\sqrt{s_{ee}}=500$ GeV due to  the exchange of excited
spin--$\frac{1}{2}$ fermions, and compare with deviations arising
from anomalous trilinear couplings between the gauge bosons. We
make a detailed study of the experimental signatures of excited
fermions and anomalous couplings, exploiting the possibility of
polarizing both the electron and laser beams and we design the
best strategy to identify the origin of the signal.  Our results
show that this reaction furnishes stronger limits than the standard
reaction $e\gamma \rightarrow e \gamma$ for excited electrons above
the kinematical limit. Furthermore, even when
the excited electron does not couple to photons, and therefore
cannot be produce in the $s$--channel, the existence of the
corresponding  excited neutrino can be detected in the single $W$
production, via its $t$--channel contribution. We also show how
the use of polarization allows the discrimination between the
excited lepton signal and the one due to the anomalous trilinear
gauge couplings.

The outline of this paper is the following. In Sec.\ \ref{lag},
we introduce the effective Lagrangians describing the excited
fermion and anomalous gauge bosons couplings, and in Sec.\
\ref{lum} we present the main ingredients of  the laser
backscattering mechanism with polarized beams. Section
\ref{results} contains the analysis of the reaction $e\gamma
\rightarrow W \nu_e$ and displays our results. Finally, in Sec.\
\ref{conclusions}, we summarize our conclusions.


\section{Effective Lagrangians}
\label{lag}

In order to describe the effects of new physics due both to the
presence of new excited leptons and to anomalous trilinear
couplings we make use of the effective Lagrangian approach. For
the excited states, we concentrate here in a specific model
\cite{hag}, which has been extensively used by several
experimental collaborations \cite{lep,hera}. In doing so, we
introduce the weak doublets, for the usual left--handed fermion
($\psi_L$) and for the excited fermions ($\Psi^\ast $), and we
write  the most general dimension-five effective Lagrangian
describing the coupling of the excited fermions to the usual
fermions, which is  $SU(2) \times U(1)$ invariant and CP
conserving,
\begin{equation} {\cal L}_{Ff} = \frac{-1}{2
\Lambda} {\bar \Psi^\ast} \sigma^{\mu\nu} \left(g f \frac{\tau^i}{2}
W^i_{\mu\nu} +  g^\prime f^{\prime} \frac{Y}{2} B_{\mu\nu}\right) \psi_L  +
\mbox{h. c.} , 
\label{l:eu:0} 
\end{equation} 
where $f$ and $f^{\prime}$ are weight factors associated to the
$SU(2)$ and $U(1)$ coupling constants ($g$ and $g^\prime$), with
$\Lambda$ being the compositeness scale, and  $\sigma_{\mu\nu} =
(i/2)[\gamma_\mu, \gamma_\nu]$.   We will assume a pure
left--handed structure for these couplings in order to comply with
the strong bounds coming from the measurement of the anomalous
magnetic  moment of leptons \cite{g-2}.

In terms of the physical fields, the Lagrangian (\ref{l:eu:0})
becomes
\begin{equation}
{\cal L}_{Ff} = - \sum_{V=\gamma,Z,W} 
C_{VFf} \bar{F} \sigma^{\mu\nu} (1 - \gamma_5) f \partial_\mu V_\nu 
 + \mbox{h. c.} \; ,
\label{l:eu}
\end{equation}
where $F = N, E$ are the excited leptons, $f = \nu, e$, and
$C_{VFf}$ is the coupling of the vector boson with the different
fermions,
\begin{eqnarray}
C_{\gamma E e} =  - \frac{\displaystyle e}{\displaystyle 4 \Lambda} 
(f + f^{\prime}) \; & , & \; \; 
C_{Z E e} =  
- \frac{\displaystyle e}{\displaystyle 4 \Lambda} 
(f \cot\theta_W - f^{\prime} \tan\theta_W) \; ,
\nonumber \\
C_{\gamma N \nu} =  
\frac{\displaystyle e}{\displaystyle 4 \Lambda} (f - f^{\prime}) \;
& , & \; \; 
C_{Z N \nu} =  
\frac{\displaystyle e}{\displaystyle 4 \Lambda} 
(f \cot\theta_W + f^{\prime} \tan\theta_W) \; ,
\nonumber \\
C_{W E \nu} =  C_{W N e} &=& 
\frac{\displaystyle e}{\displaystyle 2 \sqrt{2} \sin\theta_W \Lambda} f \; ,
\label{CV}
\end{eqnarray}
where $\theta_W$ is the weak angle with $\tan\theta_W=g^\prime/g$.

For the trilinear gauge coupling, we also write the most general
$C$ and $P$ conserving interaction Lagrangian between the charged
gauge bosons and the photon, that is $U(1)_{\text{em}}$ invariant
\cite{anoc:lag}
\begin{equation}
{\cal L}=-ie\left[ g_1^\gamma (W^\dagger_{\mu\nu} W^\mu A^\nu 
       - W^\dagger_\mu W^{\mu\nu} A^\nu) 
 + \kappa_\gamma W^\dagger_\mu W_\nu F^{\mu\nu} + 
 \frac{\displaystyle \lambda_\gamma}{\displaystyle M^2_W} 
W^\dagger_{\tau\mu} W_\nu^\mu F^{\nu\tau} \right] \; .
\label{lag:anoc}
\end{equation}

For on--shell photons, $g_1^\gamma=1$ is fixed by electromagnetic
gauge invariance since it determines the $W$ electric charge. The
coefficients $\kappa(\lambda)$ assumes the values $1(0)$ in the
SM, and are related to the magnetic moment, $\mu_W$, and the
electric quadrupole moment, ${\cal Q}_W$, of the $W$ boson,
according to
\[
\mu_W = \frac{e}{2M_W}(1+\kappa_\gamma+\lambda_\gamma)
\quad \mbox{and} \quad
{\cal Q}_W = -\frac{e}{M^2_W}(\kappa_\gamma-\lambda_\gamma) \; .
\]

\begin{figure}
\begin{center}
\mbox{\epsfig{file=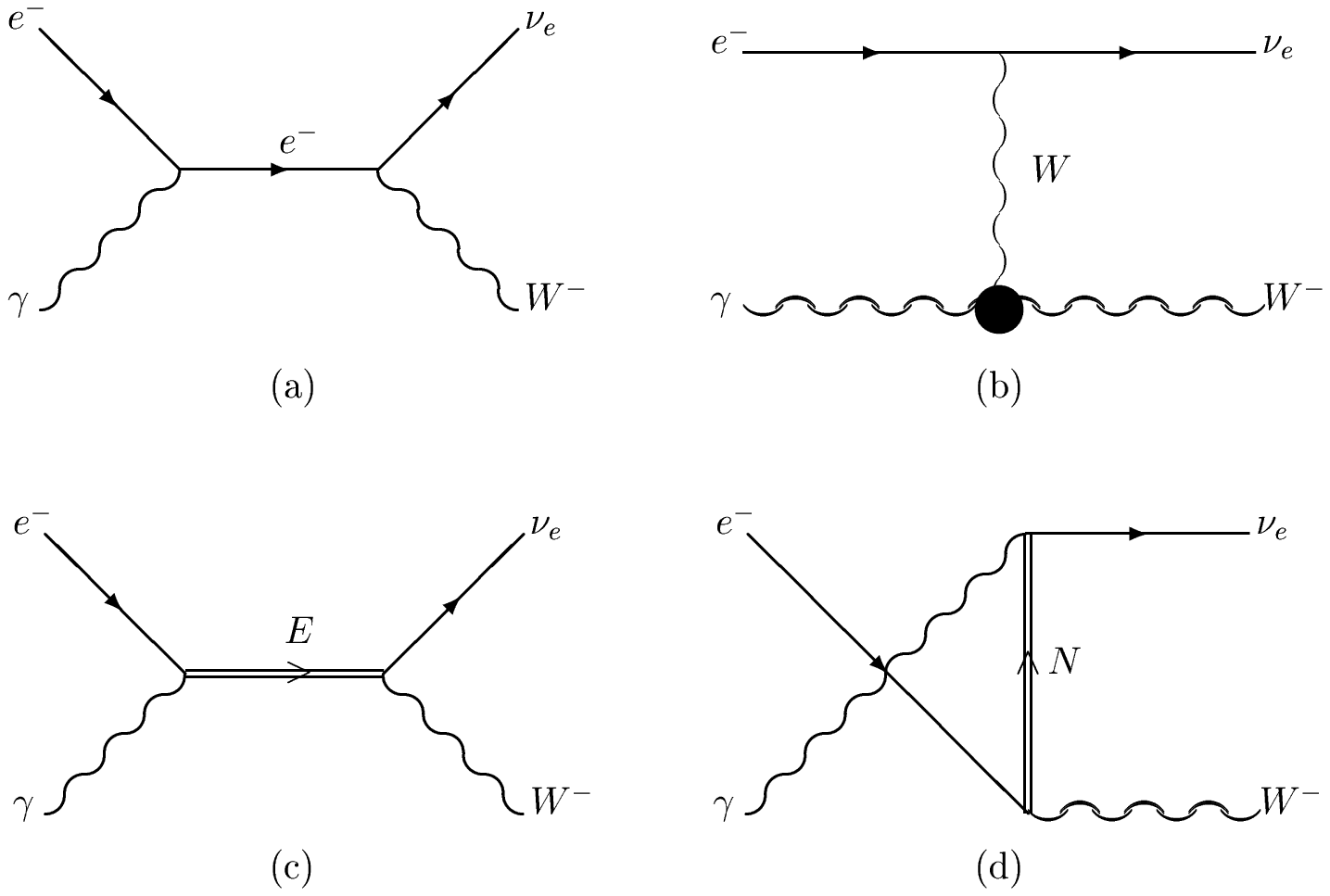,width=0.9\textwidth}}
\end{center}
\caption{Feynman diagrams contributing to the process 
$e\gamma\rightarrow W\nu_e$ in presence of excited fermions (double
lines) and anomalous vector boson coupling (black dot).}
\label{diagrams}
\end{figure}

In this paper, we are interested in analyzing the influence of both
excited fermion and anomalous vector boson coupling in the
reaction $e\gamma \to W \nu_e$. The contributions of these new
particles and interactions are represented in Fig.\ \ref{diagrams}
as double lines and black dot, respectively.

\section{Polarized Laser Backscattering}
\label{lum}

The electron beam of a linear collider can be transformed  into a
intense photon  beam via the process of laser backscattering
\cite{las0}. The basis of this mechanism lay on the fact that
Compton scattering of energetic electrons by soft laser photons
results into high energy photons, collimated in the direction of
the incident electron.  

\begin{figure}
\begin{center}
\mbox{\epsfig{file=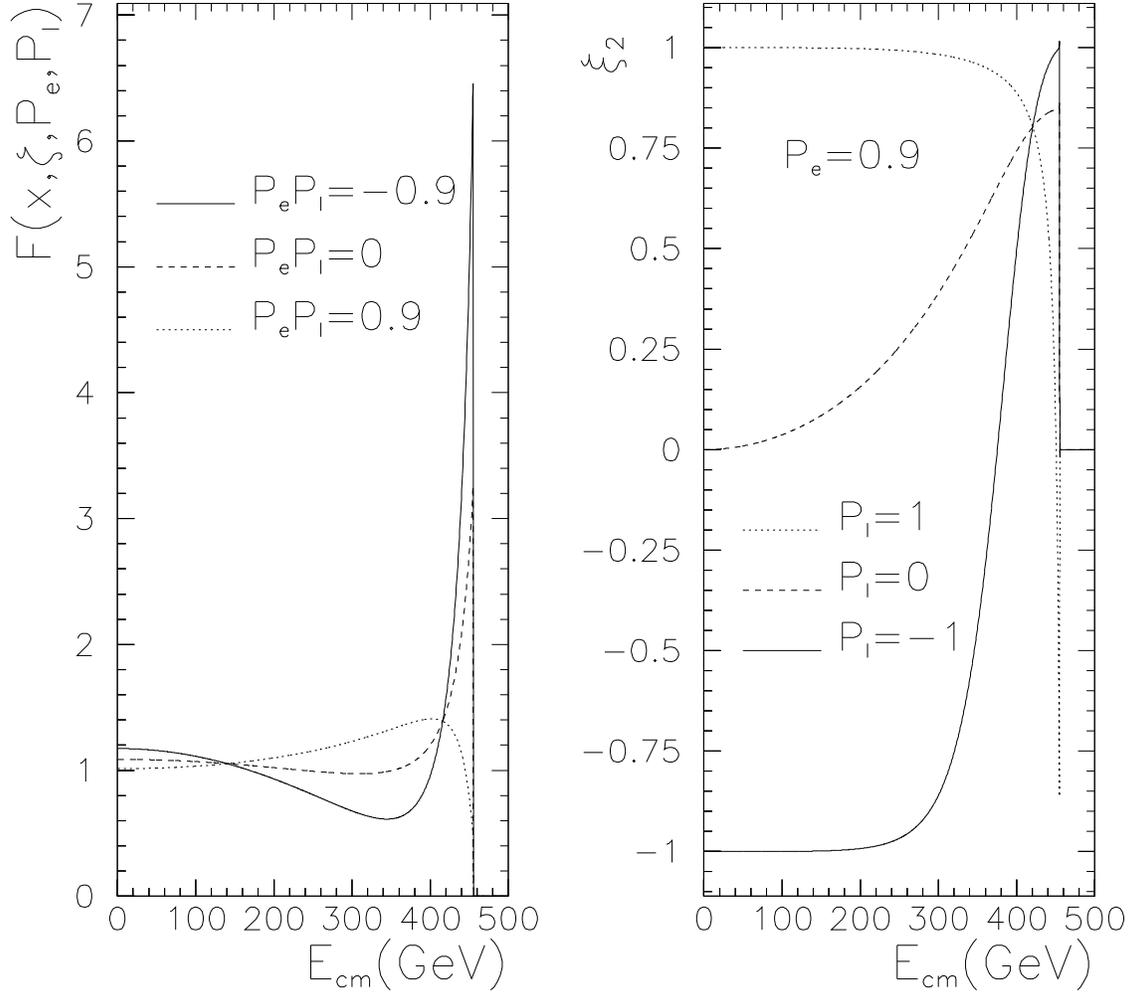,width=0.9\textwidth}}
\end{center}
\caption{Photon energy distribution and the circular polarization
distribution as function of the subprocess energy, for different
polarization configurations of the electron and laser photon.}
\label{fig:lback}
\end{figure}

The backscattered photon distribution function for polarized
electron and laser beams can be written \cite{laser:pol} as
\begin{equation}
F(x,\zeta; P_e, P_l)=\frac{2\pi \alpha ^2}{\zeta\,m^2\,\sigma_c}
\left[\frac{1}{1-x}+1-x-4r( 1- r ) - P_e P_l \; r \; 
\zeta \; ( 2r-1)(2-x ) \right] \; ,
\label{f:pol}
\end{equation}
where $P_{e}$ is the mean electron longitudinal polarization, $P_l$
is the laser photon circular polarization, and $\sigma _c$ is the
Compton cross section, which can be written as,
\begin{equation}
\sigma _c = \sigma_c^0+ P_e P_l \; \sigma_c^1  \; ,
\label{sig:com}
\end{equation}
with
\begin{eqnarray}
\sigma_c^0 &=& \frac{2\pi \alpha ^2}{\zeta\,m^2}
\left[ \left( 1- \frac{4}{\zeta} - \frac{8}{\zeta^2} \right) 
\ln \left( \zeta+1 \right) + \frac{1}{2}  + \frac{8}{\zeta} -
\frac{1}{2\left( \zeta+1\right) ^2} \right] \; ,
\nonumber \\
\sigma_c^1 &=& \frac{2\pi \alpha ^2}{\zeta\,m^2}
\left[\left( 1+\frac 2\zeta\right) \ln \left( \zeta+1\right) -
\frac{5}{2} + \frac{1}{\zeta+1} - 
\frac{1}{2\left( \zeta+1\right) ^2}
\right] \; .
\label{sig:01}
\end{eqnarray}
We have defined the variables
\begin{equation}
x = \frac{\omega}{ E} \; , 
\;\;\;\;\;\;\;\;
\zeta =\frac{4 E \omega_0}{m^2}  \; ,
\;\;\;\;\;\;\;\;
r = \frac{x}{\zeta (1-x)} \; ,
\end{equation}
where $m$ and $E$ are the electron mass and energy, $\omega_0$ is
the laser energy and $\omega$ is the backscattered photon energy.
The variable $x \leq x_{max} \equiv \zeta/(\zeta +1)$ represents
the fraction of the electron energy carried by the backscattered
photon.  A cut--off value $\zeta= 2(1+\sqrt{2}) \simeq 4.83$ is
assumed in order to avoid the threshold for  electron--positron
pair creation through the interaction of the laser and the
backscattered photons. The backscattered photon spectrum
(\ref{f:pol}) depends only on the product $P_e P_l$, and as can
be seen from Fig.\ \ref{fig:lback}a, for negative values of this
product the spectrum is dominated by hard photons. 

A very powerful feature of the Compton backscattering mechanism
is the possibility of obtaining a high degree of polarization for
the backscattered photons by polarizing the incoming electron
and the laser beams. The mean backscattered photon helicity
is given by the Stokes parameter
\begin{equation}
\xi_2 = \frac{1}{D} \left\{ P_e \,r\,\zeta\left[ 1+\left( 1-x\right) 
\left( 2r-1\right) ^2\right]  -P_l\left( 2r-1\right) 
\left[ \frac{1}{1-x} + 1 - x \right] \right\}
\label{xi:2}
\end{equation}
where
\begin{equation}
D= \frac{1}{1-x}+1-x-4r\left( 1-r\right)
- P_e P_l\,r\,\zeta\left( 2r-1\right) \left( 2-x\right)  \; .
\end{equation}

For $x = x_{max}$ (or $r=1$) and $P_e = 0$ or $P_l = \pm 1$, we
have $\xi_2 = - P_l$, {\it i.e.\/} the polarization of the
backscattered photon beam has the opposite value of the laser
polarization. One can also see that for $x = \zeta / (\zeta +2)$
(or $r=1/2$) the Stokes parameter $\xi_2$ is independent of the
laser polarization (see Fig.\ \ref{fig:lback}b), and is given by
\begin{equation}
\xi_2^{(r=1/2)} = P_e \frac{\zeta (\zeta +2)}{\zeta (\zeta +2) +4} \; .
\label{xi:2:r}
\end{equation} 

The cross section for the reaction $\gamma e \rightarrow X$ in an
electron--positron linear collider where the positron beam, with
longitudinal polarization $P_p$,  is converted into a
backscattered photon beam, is given by
\begin{equation}
d\sigma_{P_e \xi_2} \left(\gamma e \rightarrow X \right) = 
\kappa \int_{ x_{min}}^{x_{max}} dx 
\; F(x,\zeta; P_p, P_l) \; 
d\hat{\sigma}_{P_e \xi_2} (e \gamma \rightarrow X) \; ,
\label{cross:pol}
\end{equation}
where $\kappa$ is the efficiency of the laser conversion of the
electrons into photons and $d\hat{\sigma}_{P_e \xi_2}$ is the
polarized cross section for the subprocess $\gamma e \rightarrow
X$, which is a function of $\hat{s} = x s$. In our calculation we
assume that 100 \% of the electrons are converted into photons
($\kappa$=1).

The polarized subprocess cross section can be written as
\begin{eqnarray}
d\hat{\sigma}_{P_e \xi_2} &=&  \frac{1}{4}  \left[(1+P_e \xi_2 )
\left(d\hat{\sigma}_ {++} + d\hat{\sigma}_ {--} \right) 
+  (P_e + \xi_2 ) \left(d\hat{\sigma}_ {++} - d\hat{\sigma}_ {--} \right)
\right. \nonumber \\ 
&&  \left. +   (1-P_e \xi_2 )\left(d\hat{\sigma}_ {+-} + d\hat{\sigma}_
{-+}\right)  
 +  (P_e - \xi_2  )\left(d\hat{\sigma}_ {+-} -
d\hat{\sigma}_{-+}\right)  \right] \; ,
\label{elem:pol}
\end{eqnarray}
with $d\hat{\sigma}_{\lambda_e \lambda_{\gamma}}$
($\lambda_{e(\gamma)} = \pm 1$) being the polarized subprocess
cross section for full electron and photon polarization, $P_e$ is
the  longitudinal polarization of the electron beam, and the
Stokes parameter, $\xi_2$,  is given in Eq.\ (\ref{xi:2}). 

\section{Results}
\label{results}
\subsection{Excited Fermions Signature}
The standard mechanism to establish the existence of an excited
electron with mass below the kinematical reach of the $e\gamma$
machine is the identification of the Breit--Wigner profile in the
$e\gamma$ invariant mass distribution of the process
$e\gamma\rightarrow e\gamma$ \cite{e:gam}. This is obviously only
possible when the excited electron couples to the photon, i.e, $f
\ne -f^\prime$ [see Eq.\ (\ref{CV})].  On the other hand, the
reaction $e\gamma \rightarrow W\nu_e$ is sensitive both to the
exchange of the excited electron in the $s$--channel and to the
exchange of the excited neutrino in the $t$--channel (see Fig.\
\ref{diagrams}). The characteristic signature of the excited
leptons will therefore depend on the excited fermion mass and on
the relative weight of the $s$--channel versus the $t$--channel
contribution or, in other words, the relative sizes of $f$ and
$f^\prime$. 

\begin{figure}
\begin{center}
\mbox{\epsfig{file=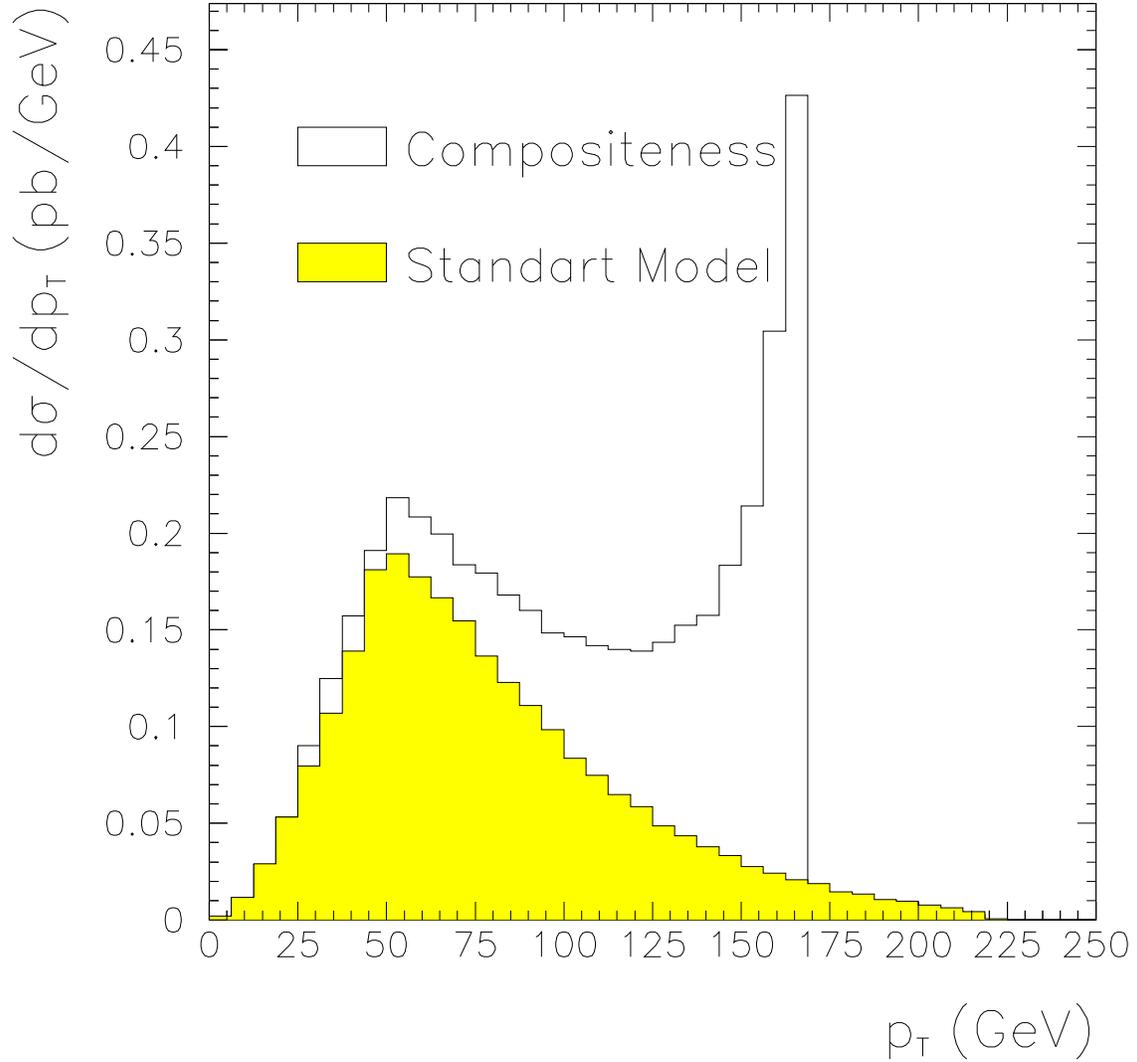,width=0.9\textwidth}}
\end{center}
\caption{Tranverse momentum distribution of $W$ bosons in
presence of an  excited electron with $M_E=350$ GeV, compared to
the SM prediction, for $f=f^\prime=1$, and $\Lambda=1$ TeV.}
\label{fig:dsig:dpt}
\end{figure}

As in the reaction $e\gamma\rightarrow e\gamma$, the existence of
an excited  electron with nonvanishing coupling to the photon and
with mass below the kinematical reach of the $e\gamma$ collider,
can also be easily identified in the reaction $e\gamma\rightarrow
W\nu_e$ but now through the study of the $W$ transverse momentum
distribution, $d\sigma/dp_T$.  For illustration, we
present in Fig.\ \ref{fig:dsig:dpt}  the $p_T$ distribution of
$W$ produced in the reaction $e\gamma\rightarrow W\nu_e$, for 
$M_E = 350$ GeV at an $e^+e^-$ collider with $\sqrt{s} = 500$
GeV.  We introduced a cut in the polar angle ($\theta$) of the
detectable final state particles with the beam pipe of $15^\circ$
to ensure their detection. We also assumed a reconstruction
efficiency of 60\%  for the produced $W$'s. As expected, the
existence of excited states with mass below the kinematical reach
of the $e\gamma$ collider provides an very clear signal, the
jacobian peak at $p_T \sim M_E/2$. This situation will be no
longer considered here, since our main concern is the
possibility of misidentification of new physics effects on the
NLC, and in this scenario, the existence of excited fermions can
easily be set apart from the anomalous vector boson contribution.

\begin{figure}
\begin{center}
\mbox{\epsfig{file=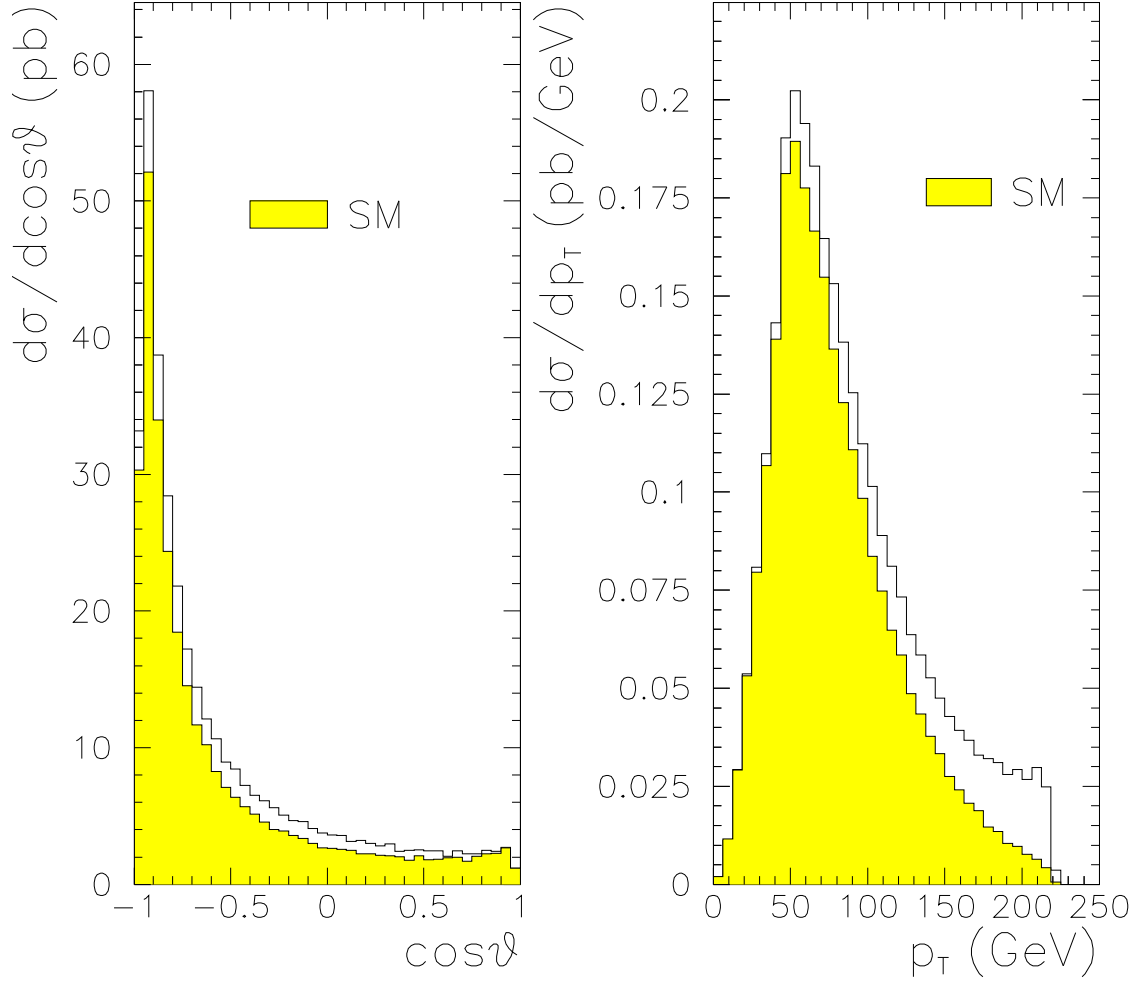,width=0.9\textwidth}}
\end{center}
\caption{$W$ boson kinematical distributions in the presence of
composite states, compared to the SM predictions, for
$f=f^\prime=1$, and  $M_E=\Lambda=500$ GeV.}
\label{fig:ex_dist}
\end{figure}

For excited leptons above the kinematical limit of the collider
we still find an enhancement on the total cross section due to
the $s$--channel contribution. We also find an effect in the
distribution of the produced  $W$.  We simulated these
distributions for $M_E=\Lambda=500$ GeV  and $f=f^\prime$.  We
present on Fig.\ \ref{fig:ex_dist} the transverse momentum
distribution, and the angle between the $W$ and the electron beam
direction for unpolarized beams. As we can see, the existence of
excited fermions would lead to an enhancement of $W$ production
at large $p_T$, which reflects the tail of the jacobian
peak.  In the angular distribution of the $W$  the effect of
composition is very small.

\begin{figure}
\begin{center}
\mbox{\epsfig{file=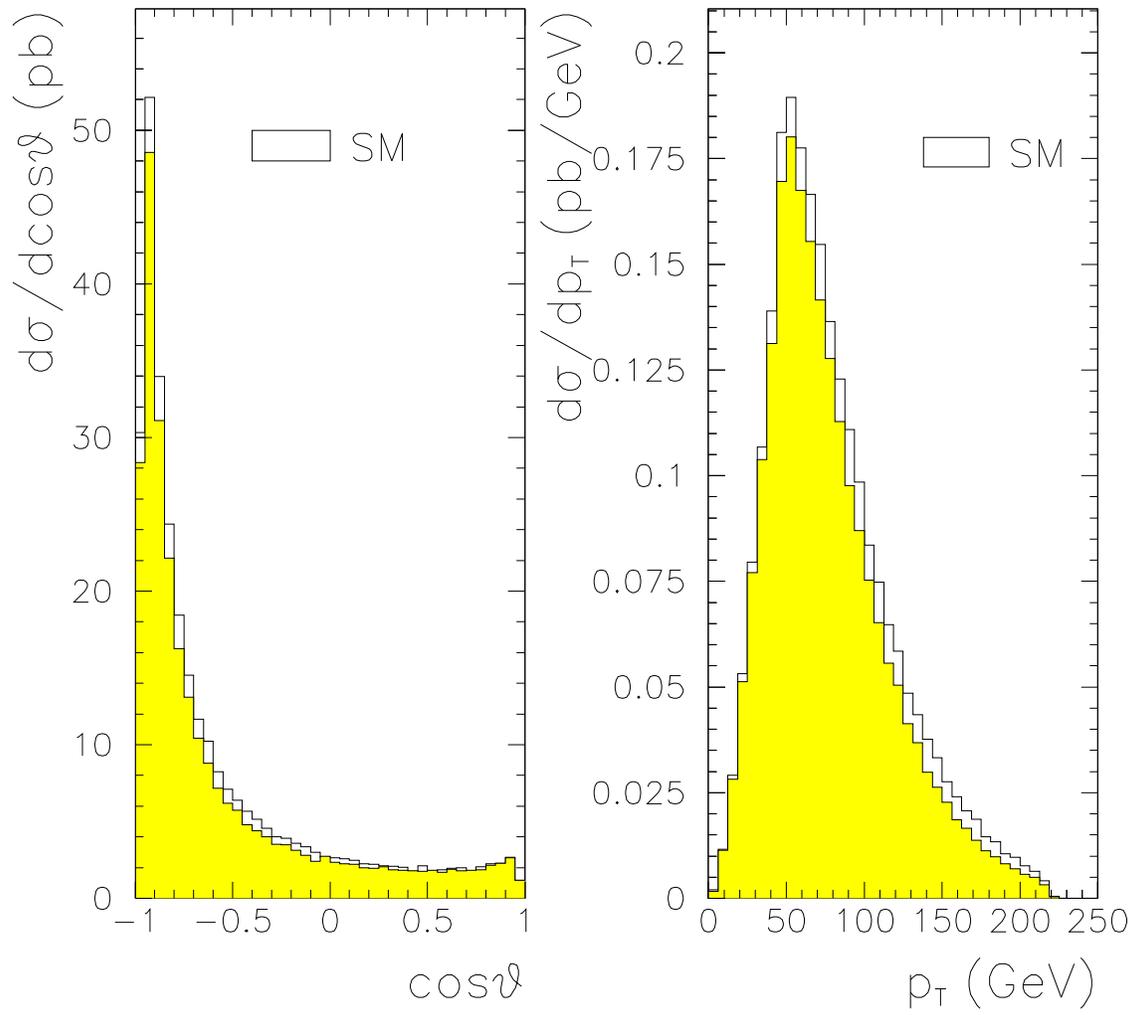,width=0.9\textwidth}}
\end{center}
\caption{The same as Fig.\ \ref{fig:ex_dist}, for
$f=f^\prime=-1$, and $M_N=\Lambda=300$ GeV.}
\label{fig:exn_dist}
\end{figure}

The process mediated by the exchange in the $t$--channel of
excited neutrinos coupled to photons, takes place when $f\ne
f^\prime$, and gives a much smaller effect. This process is
particularly interesting in the extreme situation when
$f=-f^{\prime}$ since, in this case, $C_{\gamma Ee}$ vanishes,
and just the $t$--channel diagram contributes. This contribution
gives rise to a destructive interference which diminishes the total $W$
yield, without altering the shape of the transverse momentum
and angular distributions in a significant way (see 
Fig.\ \ref{fig:exn_dist}).

In order to quantify the reach of an $e\gamma$ collider to search
for new physics, we defined the statistical significance of the
signal (${\cal S}$) as
\begin{equation} 
{\cal S} \equiv \frac{| \sigma_{\text{new}} -
\sigma_{\text{SM}}|} {\sqrt{\sigma_{\text{SM}}}} \sqrt{{\cal
L}_{\text{ee}}} \; , 
\end{equation}
where $\sigma_{\text{new}}$ ($\sigma_{\text{SM}}$) is the total
cross section for new physics (Standard Model) contributions, and
${\cal L}_{\text{ee}}$ is the integrated luminosity of the
machine. We will assume ${\cal L}_{\text{ee}} = 50$ pb$^{-1}$ for
the NLC. We impose the polar angle of all detectable final
particles with the beam pipe to be smaller than $15^\circ$ ($|
\cos\theta | \leq 0.97$), in order to ensure that the event is
well within the detector volume. With this requirement, we
assumed that  60\% of the produced $W$'s can be properly
reconstructed. 

\begin{figure}
\begin{center}
\mbox{\epsfig{file=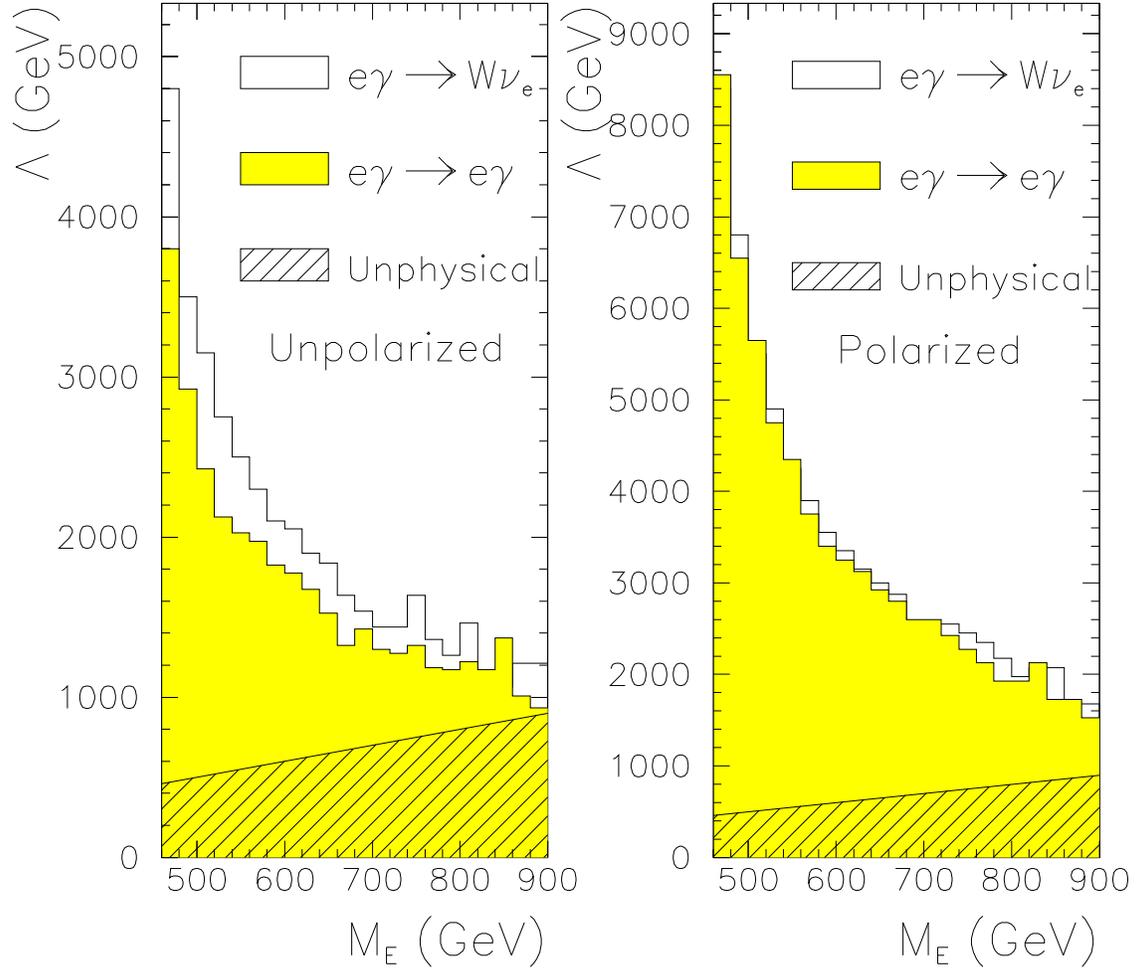,width=0.9\textwidth}}
\end{center}
\caption{Discovery contour for $f=f^\prime=1$.
Inside the shaded regions the deviations from SM are greater
3$\sigma$. We excluded the unphysical region where $\Lambda <
M_E$. }
\label{fig:disc}
\end{figure}

\begin{figure}
\begin{center}
\mbox{\epsfig{file=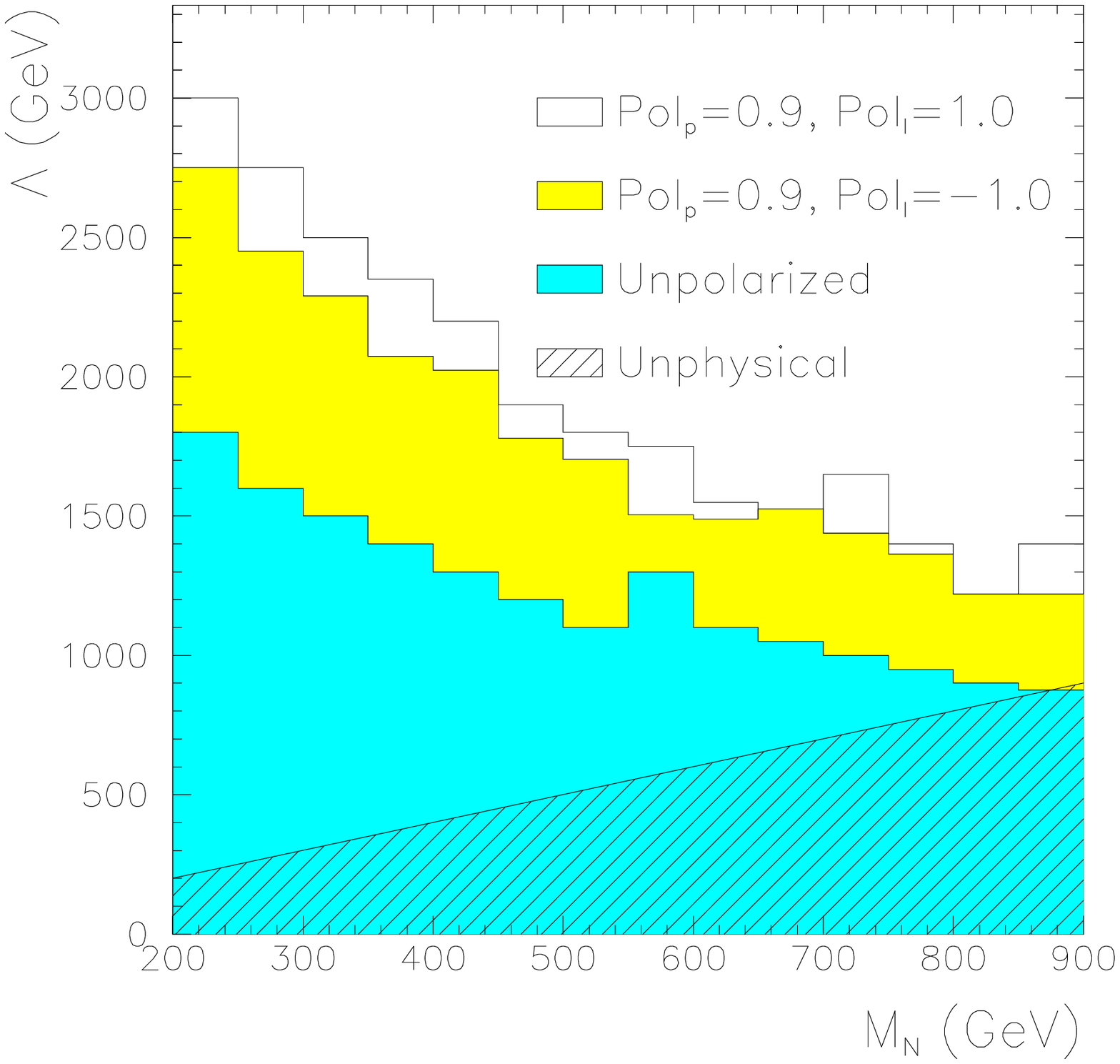,width=0.9\textwidth}}
\end{center}
\caption{Discovery contour for composite neutrinos, for
$f=-f^{\prime}$. Inside the shaded regions, deviations from SM
are greater then 3$\sigma$. We excluded the unphysical region
where $\Lambda < M_N$.}
\label{fig:discn}
\end{figure}

In Fig.\ \ref{fig:disc}, we show the discovery limits for the
composite state in the $\Lambda\times M_E$ plane for both
$e\gamma$ and $W\nu_e$ final states, requiring a 3$\sigma$ effect
in the total cross. As we can see, for unpolarized photons, the
$e W$ channel  furnishes stronger limits than the $e\gamma$ channel
which suffers from large background from Compton scattering.
However an excess in the $e\gamma$ production is a definite
signal of compositeness, since there is no contribution
of anomalous couplings in this channel. Moreover, as mentioned
above, if the weight factors $f$ and $f^\prime$ are such  that
$f=-f^\prime$, the signal of composition on the $e\gamma$
production will disappear completely, but survive on single $W$
production through the exchange of a neutral composite fermion on
the $t$--channel. The observability limit of the NLC for such
neutral states is shown in  Fig.\ \ref{fig:discn}.

Polarization can be used to improve the discovery region in the
$\Lambda\times M$ plane through the enhancement of the luminosity
and the cross section.  The photon  distribution functions assume
approximately the same value at $\bar{x} = \zeta / (\zeta +2)
\simeq 0.71$, even for different polarization configurations of
the initial particles (see Fig.\ {fig:lback}). In the interval
$0< x < \bar{x}$, the luminosity is higher for $P_p P_l > 0$,
whereas for the range $x > \bar{x}$ the distribution with $P_p
P_l < 0$ dominates. Therefore, in order to search for excited
electrons with mass above $\overline{M_E}=\sqrt{\bar{x}s}$,  we
should employ the polarization configurations of the positron and
the laser in such a way that $P_p P_l < 0$.  The degree of
circular polarization of the scattered photon, $\xi_2$, has the
same sign as the positron polarization in the region of interest.
Because of the chiral nature of  the weak interactions, just
left-hand electrons will produce $W$'s, and therefore, only
electrons and photons with negative helicity contribute to the
exchange of an excited electron in the $s$--channel. In
consequence, the best choice of the polarization parameters for
this case is to set  electron and positron beams with negative
polarization, while  the laser is set with positive polarization.
We assume that a 90\% of beam polarization is achievable at the
NLC  and that the laser can be 100\% polarized ({\it i.e.\/}
$P_{e^-}=P_{e^+}=-0.9$, and $P_l=1$). With this setup the
discovery region can be enlarged as much as shown on Fig.\
\ref{fig:disc}.

Things change when we are dealing with the process involving the
excited neutrino. As before, just left--handed electrons will
participate in the reaction and again one must choose
$P_{e^-}<0$, but now the photon  line is attached to the final
state neutrino, and consequently, just the  positive helicity
photons couple to the fermionic line. According to the 
discussion above, to obtain a highly energetic and positively polarized
photon beam, we must set the positron beam with positive
polarization ($P_{e^+}=0.9$) and the laser with negative
polarization ($P_{l}=-1$). However, one must also notice that in
this process, by setting the positron and laser with positive
polarization, one achieves to have most of the photons with
positive helicity (except the very high energy ones) while
reducing the SM background from the diagram with self boson interaction. 
Due to this reduction in the SM background,
this configuration with $P_{e^-}=-P_{e^+}=-0.9$ and $P_l=1$
proves to furnish better limits as can be seen in Fig.\
\ref{fig:discn}.
 
\subsection{Anomalous Gauge Boson Couplings: Comparison}
\begin{figure}
\begin{center}
\mbox{\epsfig{file=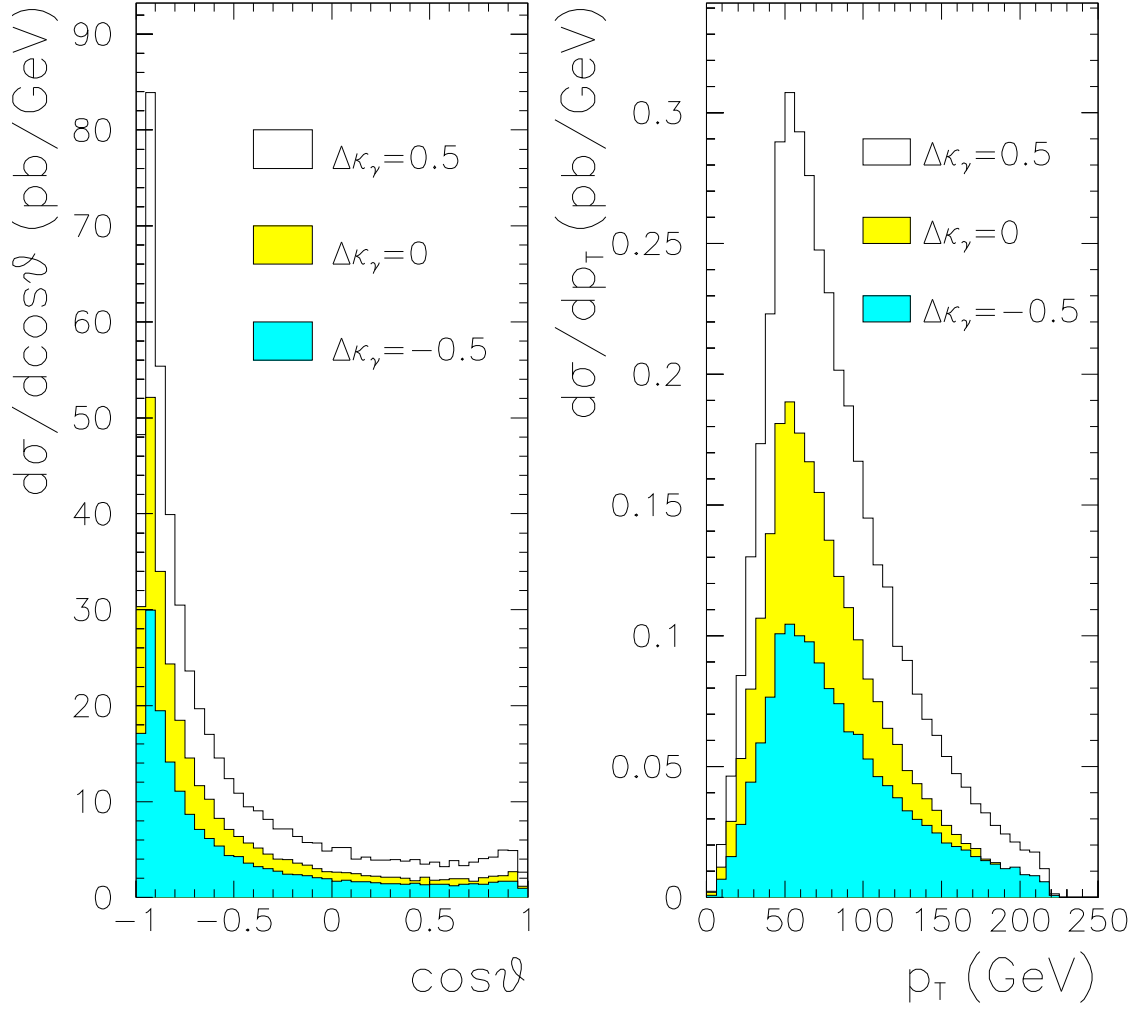,width=0.9\textwidth}}
\end{center}
\caption{Effects of a variation on $\Delta\kappa_\gamma$ in $W$
kinematical  distributions, compared with the Standard Model
($\Delta\kappa_\gamma = 0$).}
\label{fig:ack_dist}
\end{figure}

Let us now examine the consequences of the existence of trilinear
anomalous couplings in the single $W$ production. In order to
make clear the effect of each of the two possible anomalous
couplings ($\Delta\kappa_\gamma$ and $\lambda_\gamma$), we envisaged
two distinct scenarios, by keeping just one of them different
from zero at a time. In Fig.\ \ref{fig:ack_dist}, we show the
angular and transverse momentum distributions for different
values of $\Delta \kappa_\gamma = 1-\kappa_\gamma$ 
while keeping $\lambda_\gamma = 0$. We can see from these figures
that the only effect of varying $\Delta\kappa_\gamma$ is to
increase or decrease the total cross section depending on its
sign, while producing  small effect on the shape of the
distributions. The number of observed events is larger (smaller)
than the SM expectation  for $\Delta \kappa_\gamma > 0$
($\Delta \kappa_\gamma < 0$).

\begin{figure}
\begin{center}
\mbox{\epsfig{file=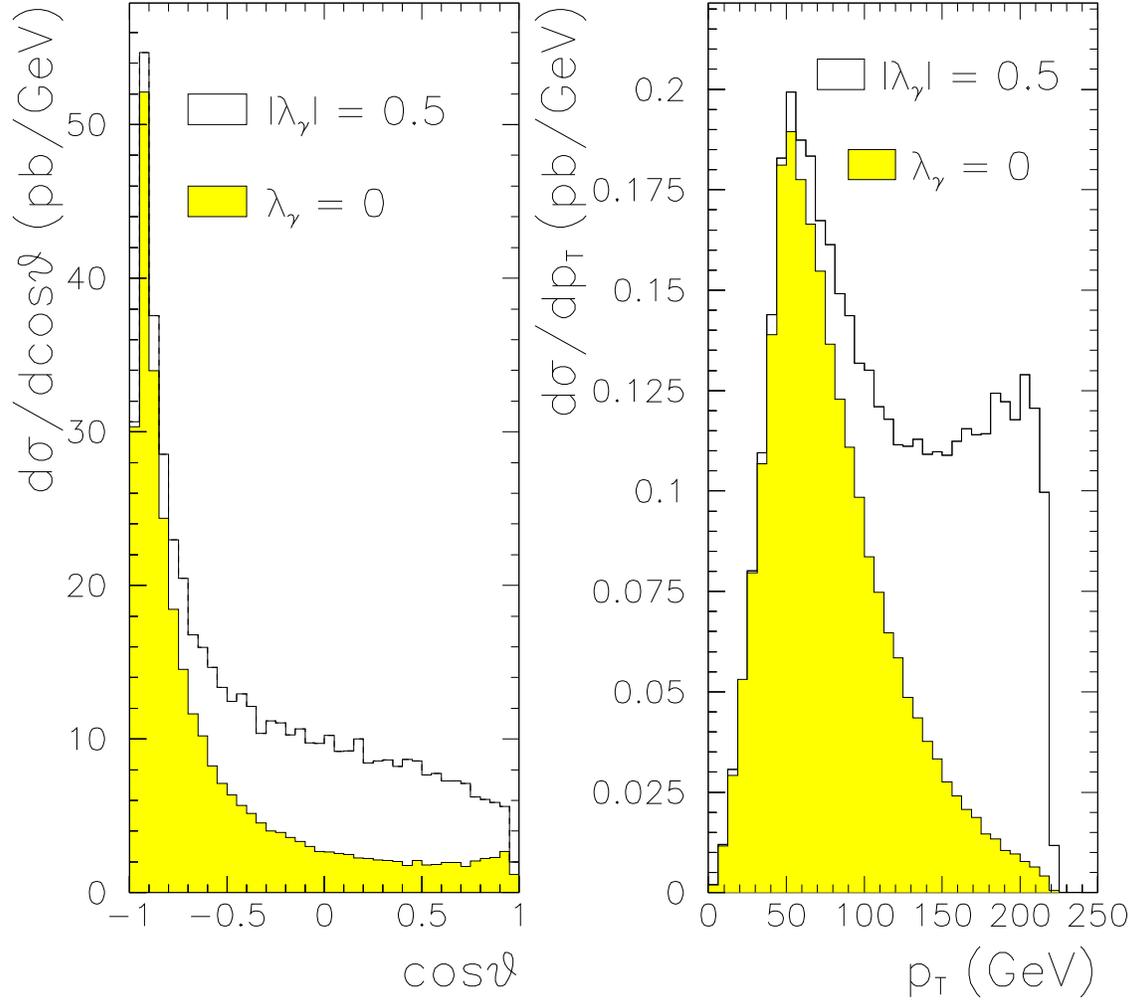,width=0.9\textwidth}}
\end{center}
\caption{Effects of a variation on $\lambda_\gamma$ in the $W$
distributions, compared with the Standard Model ($\lambda_\gamma = 0$).}
\label{fig:acl_dist}
\end{figure}

Conversely, in Fig.\ \ref{fig:acl_dist}, we show the angular and
transverse momentum distributions for varying  $\lambda_\gamma$
while keeping $\Delta\kappa_\gamma = 0$. We can see in these
figures that the cross section is always larger than  the SM
prediction for any sign of the coupling  $\lambda_\gamma$. The
presence of a nonvanishing $\lambda_\gamma$ also increases  the
number of $W$'s produced in the central and forward direction and
of those produced with high $p_T$.

Polarization has also proven to be very useful to unravel the
existence of anomalous couplings and the discovery 
limits for anomalous couplings have been extensively covered elsewhere
\cite{anoc:pheno}. In what follows we will concentrate on the
possibility of distinguishing these effects from those arising
from the existence of the excited leptons.

Let us now figure out the different scenarios we can face when
the NLC starts its operation. It will probably start on its
unpolarized mode, in order to be as ``democratic'' as possible to
discover new physics. The first observable that the accumulated
statistics will permit to measure is the total cross section.
From what we saw above, the effect of new physics  can either
increase or decrease the total $W$ production.

If the number of produced $W$'s is greater than expected, one
should just look into the $e\gamma$ events that will be produced
at  the same time. As we saw, in the framework where composition
leads to an enhancement on the total single $W$ cross section,
the $e\gamma$ channel is also very sensitive  to the presence of
excited leptons. An increase in the  $e\gamma$ cross section
would indicate that the excess of $W$ is due to the existence of
an excited electron.  If no excess is seen in the $e\gamma$
channel, likelihood fittings to the angular and transverse
momentum distributions will be able to determine the anomalous
coupling parameters leading to such deviations.

On the other hand, if the total cross section is smaller than
predicted by the SM, two possibilities remain:  (i) the trilinear
coupling is anomalous with  the value of $\Delta\kappa_\gamma\leq
0$, or, (ii) the excited neutrino exists, and its gauge structure
is such that $f \simeq -f^{\prime}$. To distinguish these two
possibilities, we define the polarization asymmetry
\begin{equation}
A_{\text{pol}}= \frac{\Delta \sigma_{+-}-\Delta \sigma_{-+}}
              {\Delta \sigma_{+-}+\Delta \sigma_{-+}}
\end{equation}
where, for instance,  
\begin{equation}
\Delta \sigma_{+-}=\sigma^{\text{SM}}_{+-}- \sigma^{\text{obs}}_{+-}
\end{equation}
measures the deviation from the SM prediction when the positron
and laser polarizations are set  $P_{e^+}=0.9$ and $P_l=-1$, 
always keeping $P_{e^-}=-0.9$. Correspondingly, $\Delta
\sigma_{-+}$ quantifies the  deviation for the SM prediction for
$P_{e^+}=-0.9$ and $P_l=1$. 

\begin{figure}
\begin{center}
\mbox{\epsfig{file=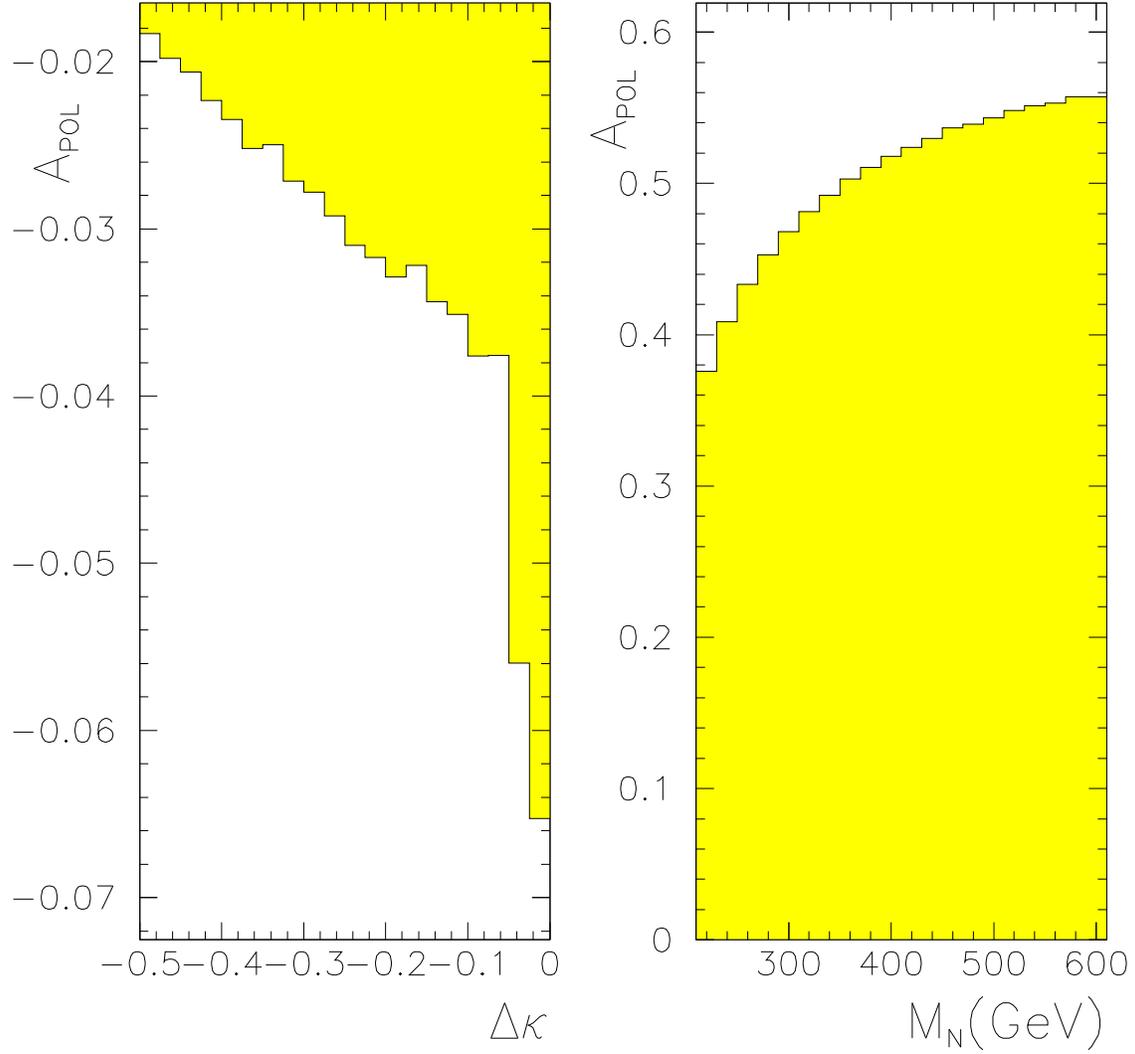,width=0.9\textwidth}}
\end{center}
\caption{Polarization asymmetry due to the presence of anomalous 
couplings with $\Delta \kappa_\gamma\leq 0$ and in models with excited 
neutrinos for $f=-f^\prime=1$, and $M_N=\Lambda$.}
\label{fig:asym}
\end{figure}

In Fig.\ \ref{fig:asym}, we plot $A_{\text{pol}}$ for the two
type of models here considered. As seen in this figure the
asymmetry due to the presence of anomalous trilinear  gauge
couplings $\lambda_\gamma=0$ and  $\Delta\kappa_\gamma \leq 0$ is
always negative and very small. On the contrary, the presence of
excited neutrinos would yield a larger positive  asymmetry.  This
is understood because photons with both polarizations contribute
to the anomalous $\Delta\kappa_\gamma$ term, while only positive
helicity photons  enter in the excited neutrino contribution.
Therefore in both configurations the reduction in the cross
section due to the negative $\Delta\kappa_\gamma$ is of the same
order and the corresponding asymmetry is small. However in the
configuration $(+-)$, since most of the high energy photons have
positive helicity, the effect of the excited neutrino
contribution is enhanced and the destructive interference is
larger. Consequently, $\Delta \sigma_{+-} \gg \Delta \sigma_{-+}$
what gives a positive and larger asymmetry.

\section{Conclusion}
\label{conclusions}

In this work, we have studied the deviations from the SM
predictions for the reactions $e\gamma \rightarrow W \nu_e$  at
$\sqrt{s_{ee}}=500$ GeV due to two possible sources of new
physics. We have analyzed the effect associated with the
existence of excited spin--$\frac{1}{2}$ fermions, and  we have
compared them to those arising from  anomalous trilinear
couplings between the gauge bosons. We have discussed how the use
of polarization can improve the reach of the machine in the
search for excited fermions. Our results show that this reaction
can furnish stronger limits than the standard reaction $e\gamma
\rightarrow e \gamma$ for excited electrons above the kinematical limit. 
We have shown how, by setting up the
electron, positron, and laser polarizations, we can be sensitive
to scales, $\Lambda$, up to  $9$ TeV, which is twice the bound
obtained just with  unpolarized beams.  Our results show that
excited electrons coupling to photons with strength $e$  can be
ruled out for $M_{E}\leq 1$ TeV.  Also, the simultaneous analysis
of both $e\gamma$ and $W \nu_e$ channels allows to discriminate
the excited electron signature from the one due to the presence
of anomalous trilinear gauge couplings which would lead to the
same increase in the total $W$ yield. 

Single $W$ production is also the main channel to look for
excited neutrinos when the corresponding excited electron does
not couple to  photons. In this case, we get a reduction in the
number of events, as compared to the SM prediction, due to the
destructive interference between the SM contribution and the one
due to the exchange of the excited neutrino in the $t$--channel.
This reduction is significative enough to rule out excited
neutrinos with $M_{N}\leq 1$ TeV. Anomalous trilinear couplings
with  very small $\lambda_\gamma$ and $\Delta\kappa_\gamma\leq 0$
would also lead to a decrease in the number of events which could
fake the existence of a excited neutrino. For this case we have
introduced a polarization asymmetry which is sensitive to the
origin  of the deviation.


\acknowledgments

M.\ C.\ Gonzalez--Garcia is grateful to the Instituto de F\'{\i}sica
Te\'orica of the Universidade  Estadual Paulista for its kind
hospitality. This work was supported by Funda\c{c}\~ao de Amparo
\`a Pesquisa do Estado de S\~ao Paulo, by DGICYT under grant
PB95-1077, by CICYT under  grant AEN96-1718,  and by
Conselho Nacional de Desenvolvimento Cient\'{\i}fico e
Tecnol\'ogico.



%
%

\end{document}